\documentclass[useAMS,usenatbib]{mn2e}
\input epsf
\usepackage[usenames,dvips]{color}
\usepackage{graphicx,natbib}
\usepackage{hyperref}
\usepackage{amsmath}
\usepackage{multirow}

\title[Swift Observations of the Be/X-ray Transient System\\ 
        1A 1118--615]{Swift Observations of the Be/X-ray Transient System 1A 1118--615}
\author[L. C.-C. Lin, J. Takata, A. K. H. Kong and C. Y. Hwang]{Lupin Chun-Che Lin$^1$\thanks{E-mail:
lupin@crab0.astr.nthu.edu.tw},  Jumpei Takata$^2$, Albert K. H. Kong$^{3,4}$ and Chorng-Yuan Hwang$^1$\\
$^1$Graduate Institute of Astronomy, National Central University, Jhongli 32001, Taiwan\\
$^2$Department of Physics, University of Hong Kong, PRC\\
$^3$Institute of Astronomy and Department of Physics, National Tsing Hua University, Hsinchu 30013, Taiwan\\
$^4$Kenda Foundation Golden Jade Fellow}

\begin{document}

\date{Feb 2010; ??? 2010}
\pagerange{\pageref{firstpage}--\pageref{lastpage}}
\pubyear{2010}
\maketitle

\label{firstpage}

\begin{abstract}
We report results of {\it Swift} observations for the high mass Be/X-ray binary system 1A 1118--615, during an outburst stage in January, 2009 and at a flaring stage in March, 2009.
Using the epoch-folding method, we successfully detected a pulsed period of 407.69(2) sec in the outburst of January and of 407.26(1) sec after the flare detection in March. 
We find that the spectral detection for the source during outburst can be described by a blackbody model with a high temperature 
($kT\sim 1-3$~keV) and a small radius ($R\sim 1$~km), indicating that the emission results from the polar cap of the neutron star. 
On the other hand, the spectra obtained after the outburst can further be described by adding an additional component with a lower temperature ($kT\sim 0.1-0.2$~keV) and a larger emission radius ($R\sim 10-500$~km), which indicates the emission from around the inner region of an accretion disk. 
We find that the thermal emission from the hot spot of the accreting neutron star dominates the radiation in outburst; the existence of both this X-ray contribution and the additional soft component suggest that the polar cap and the accretion disk emission might co-exist after the outburst.
Because the two-blackbody signature at the flaring stage is a unique feature of 1A 1118--615, our spectral results may provide a new insight to interpret the X-ray emission for the accreting neutron star. 
The time separation between the three main outbursts of this system is $\sim 17$~years and it might be related to the orbital period.
We derive and discuss the associated physical properties by assuming the elongated orbit for this specific Be/X-ray transient.   
\end{abstract}


\section{Introduction}

The transient X-ray source 1A 1118--615 was first discovered by {\it Ariel V} in outburst during an observation (1974 December to 1975 January) of Cen X-3 \citep{ESWR75}.
A pulsation of 405.3(6) sec was also detected by \citet{ISB75} during this observation but it was wrongly attributed to an orbital period with an assumption that 1A 1118--615 consists of two massive compact objects.
The optical counterpart was identified as the Be star He 3-640/WRA 793 \citep{CI75} and is classified as an O9.5 \uppercase\expandafter{\romannumeral 3}--\uppercase\expandafter{\romannumeral 5}e massive star with a distance of $5\pm 2$~kpc \citep{JIC81}.
Long-term and short timescale variability in the Balmer emissions of 1A~1118--615 has been reported by \citet{Mot88} and \citet{PVG93}, respectively .
This source is also the most extreme Be/X-ray transient system known in terms of the strength of its H$_{\alpha}$ emission \citep{Coe94}. 
\citet{PVG93} also identified a number of other lines (Fe$_{\rm{\uppercase\expandafter{\romannumeral 2}}}$, He$_{\rm{\uppercase\expandafter{\romannumeral 1}~and~\uppercase\expandafter{\romannumeral 2}}}$, Si$_{\rm{\uppercase\expandafter{\romannumeral 2}}}$, C$_{\rm{\uppercase\expandafter{\romannumeral 3}}}$, N$_{\rm{\uppercase\expandafter{\romannumeral 3}}}$ and a tentative identification of Cr$_{\rm{\uppercase\expandafter{\romannumeral 1}}}$); this indicates that 1A~1118--615 has an extremely unusual Be-type primary with complex emission and absorption features.

The orbital period of 1A 1118--615 is not well studied.
Although the X-ray flux of the source is variable, it remains in the quiescent state for most of the time. 
The Be/X-ray binary system underwent a flaring state detected by the Burst and Transient Source Experiment (BASTE) on board the Compton Gamma-Ray Observatory ({\it CGRO}) \citep{Coe94} and the WATCH experiment on board the {\it GRANAT} satellite \citep{LBC92} in the late January of 1992.
More recently, 1A 1118--615 showed an outburst that triggered the Burst Alert Telescope of {\it Swift} ({\it Swift}/BAT) on 2009 January 4 \citep{MBG2009}.
Initial X-ray data analysis was done by \citet{Kong2009} and \citet{Man2009}.
1A 1118--615 was detected flaring again during February and March of 2009 by both the Joint European X-ray Telescope (JEM-X) and the soft gamma-ray imager (ISGRI) instruments on the International Gamma-Ray Astrophysics Laboratory ({\it INTEGRAL}) \citep{LWL2009}.
The X-Ray Telescope of {\it Swift} ({\it Swift}/XRT) also made follow-up observations after the flaring in March.
We here report the {\it Swift} observations of 1A 1118--615 in January and March of 2009.

\begin{table}

\caption[Swift observations list]{\footnotesize{Log of Swift observations for the PC and WT modes}}
\resizebox{1.05\columnwidth}{!} {
\begin{tabular}{clccc}
 \hline
ObsID & Epoch (UT) & Exposures (s) & Pulsed counts & Total counts$^{a}$ \\
 &  &  (PC/WT) & (WT) & (PC/WT) \\
\hline
1 & 2009 Jan04 : 03:36 & 105/549  &  351  & ---/3865 \\
2 & 2009 Jan04 : 10:05 & ---/314 & 209 & ---/2061 \\
3 & 2009 Jan04 : 13:21 & 5729/12818  & 18697 & 5729/127740 \\
4 & 2009 Jan06 : 12:07 & 2595/1041 & 249 & 1969/5950 \\
5 & 2009 Jan07 : 10:18 & 462/2746 & 3186 & ---/35340 \\
\hline
6 & 2009 Mar04 : 22:16 & 75/4270 & 1605 & ---/9817 \\
7 & 2009 Mar08 : 04:47 & ---/5221 &  4593 & ---/24195 \\
8 & 2009 Mar08 : 15:51 & ---/3275 &  2074 & ---/13775 \\
9 & 2009 Mar14 : 00:39 & ---/5627 &  1506 & ---/7249 \\
10 & 2009 Mar14 : 16:24 & ---/4109 &  971 & ---/5177 \\
\hline
\multicolumn{5}{l}{$^{a}$ Only the photons observed within 0.3--10 keV were counted.}\\ 
\end{tabular}
\\
}
\label{Oblog}
\end{table}

\section{Observations and data analysis}

We analyzed the X-ray data observed by {\it Swift}/XRT.
These data were mainly observed within two time intervals.
From 2009 January 4, the Be/X-ray transient was monitored by {\it Swift}/XRT for 4 days with a total exposure of 26 ks, which includes 17 ks in the Windowed Timing (WT) mode and 9 ks in the Photon Counting (PC) mode during the outburst.
{\it INTEGRAL} detected flares of 1A 1118--615 again from {2009} February 3 to March 3 for a total exposure of 270 ks.
The same field was extensively observed by {\it INTEGRAL} and {\it Swift} over the following days, and our analyses focus on the observations made over an 11 day period with {\it Swift}/XRT from 2009 March 4 with a total exposure of 22.6 ks (22.5 ks in the WT mode and only 75 s in the PC mode).
Details of the observations are listed in Table~\ref{Oblog}.

All the {\it Swift}/XRT data were processed with standard procedures (xrtpipeline v.0.12.0), filtering, and screening criteria by using FTOOLS (v.6.5) in the HEAsoft package (v.6.5.1). 
We obtained a refined X-ray position of the source by the task ``xrtcentroid'' from the image of ObsID 1 (Table~\ref{Oblog}) investigated with the PC mode. 
The source located at (J2000) R.A.=$11^h20^m57^s.8$, decl.=$-61^{\circ}54'57''.9$ with an uncertainty of $3''.9$ (90\% confidence level) is the only bright X-ray object in the field \citep{Kong2009}, and the source in all the other PC images follow this position well.
We extracted the source counts from a circular region within a radius of $47''$ (20 pixels) which is consistent with the 92\% of the encircle energy function.
The background is obtained from a source-free neighborhood of our target with the same extraction size.

\begin{figure}
\centering
\includegraphics[width=6.5cm,angle=-90]{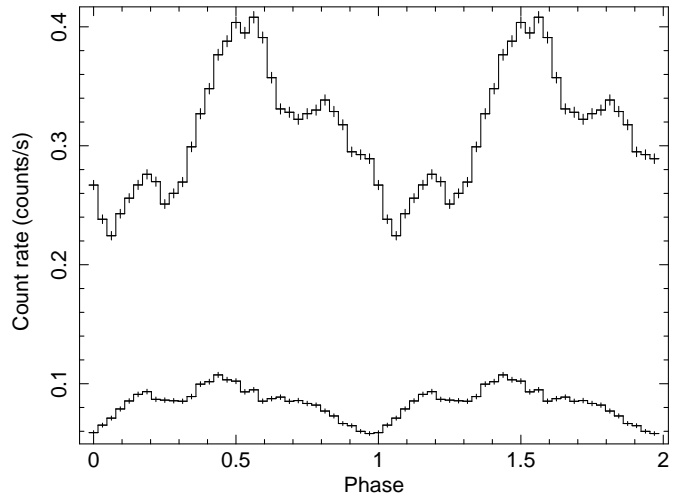}
\caption{{\small Folded 0.3--10 keV {\it Swift} light curves of 1A 1118--615. The upper curve shows the pulsation of 407.69(2) s in the January outburst with a pulsed luminosity of $5.3\times 10^{35}$ ergs/s. The lower one shows the pulsation of 407.26(1) s after the flare detection in March with a pulsed luminosity of $1.6\times 10^{35}$ ergs/s. The distance of the source is assumed to be 4 kpc \citep{CP85}.}}
\label{PsFile}
\end{figure}


\begin{table*}
\caption{\small{Best--fit spectral parameters for the non-thermal spectroscopy of 1A 1118--615. The radius is measured from the normalization factor for a source distance of 4 kpc and the flux is measured in the range 0.5--10 keV. $P_{F-test}$ is the chance probability that the improvement of the fit, compared to the single blackbody model (shown in Table~\ref{Tspresult}). }}\label{NTspresult}
\resizebox{2.15\columnwidth}{!} {
\begin{tabular}{cl|c|c|cc|cc|c} 
\hline
\multicolumn{2}{c|}{Observed Time} & Jan. 04 & Jan. 04 & Jan. 04 & Jan. 04 & Jan. 06 & Jan. 06 & Jan. 07
\\
\multicolumn{2}{c|}{/model} & 03:36(WT) & 10:05(WT) & 13:21(PC) & 13:21(WT) & 12:07(PC) & 12:07(WT) & 10:18(WT)
\\ 
\hline
\multirow{4}{*}{PO} & $N_H$ (10$^{22}$ cm$^{-2}$) & 2.40$^{+0.36}_{-0.31}$ & 1.20$^{+0.38}_{-0.31}$ & 1.40$^{+0.22}_{-0.19}$ & 1.24$^a$ & 2.51$^{+0.63}_{-0.57}$ & 2.40$^{+0.29}_{-0.27}$ & 1.08$^a$
\\ 
 & $\Gamma$ & 1.09$\pm$0.14 & 0.36$\pm$-0.17 & 0.53$^{+0.10}_{-0.09}$ & 0.66$^a$ & 0.43$^{+0.20}_{-0.18}$ & 0.93$\pm$0.11 & 0.58$^a$
\\
 & $\chi^2_{\nu}$/d.o.f. & 1.82/68 & 1.61/36 & 1.75/100 & 3.77/755 & 0.86/34 & 1.64/104 & 2.05/514
\\
 & $F_{\rm{PO}}$ ($10^{-10}$ erg cm$^{-2}$ s$^{-1}$) & 11.2 & 10.1 & 9.5 & 13.4 & 5.9 & 8.6 & 17.1
\\ 
\hline
 & $N_H$ (10$^{22}$ cm$^{-2}$) & 1.39$^{+0.44}_{-0.39}$ & 1.23$^{+0.94}_{-0.57}$ & 0.84$^{+0.65}_{-0.37}$ & 0.59$^{+0.10}_{-0.08}$ & -- & 1.33$^{+0.34}_{-0.30}$ & 0.42$^{+0.11}_{-0.07}$
\\ 
 & $\Gamma$ & 6.42$^{+1.77}_{-1.84}$ & 6.30$^{+2.09}_{-2.14}$ & 4.79$^{+5.20}_{-2.35}$ & 4.15$^{+0.36}_{-0.37}$ & -- & 6.92$^{+1.94}_{-2.53}$ & 5.18$^{+1.35}_{-0.04}$
\\
 PO & $kT_{\rm{PC}}$ (keV) & 1.65$^{+0.11}_{-0.09}$ & 2.02$^{+0.24}_{-0.21}$ & 2.10$^{+0.14}_{-0.13}$ & 1.94$^{+0.03}_{-0.02}$ & -- &  1.81$^{+0.09}_{-0.10}$ & 2.00$\pm$0.04
\\
 + & $R_{\rm{PC}}$ (km) & 1.36$^{+0.15}_{-0.13}$ & 0.99$^{+0.19}_{-0.16}$ & 0.91$^{+0.10}_{-0.08}$ & 1.21$\pm$0.02 & -- &  1.05$^{+0.10}_{-0.08}$ & 1.31$\pm$-0.04
\\
 BB & $\chi^2_{\nu}$/d.o.f. & 1.06/66 & 0.77/34 & 1.25/98 & 1.27/753 & -- & 1.18/102 & 1.12/512
\\
 & $F_{\rm{PO+BB}}$ ($10^{-10}$ erg cm$^{-2}$ s$^{-1}$) & 14.2 & 14.0 & 8.5 & 11.2 & -- & 9.6 & 14.1
\\
 & $P_{F-test}$ & 0.026 & 0.018 & 0.28 & 1.0$\times$10$^{-10}$ & -- &  0.065 & 1.3$\times$10$^{-5}$
\\
\hline
\multicolumn{2}{c|}{Observed Time} & Mar. 04 & Mar. 08 & \multicolumn{1}{c|}{Mar. 08} & Mar. 14 & \multicolumn{1}{c}{Mar. 14} & \multicolumn{2}{c}{}
\\
\multicolumn{2}{c|}{/model} & 22:16(WT) & 04:47(WT) & \multicolumn{1}{c|}{15:51(WT)} & 00:39(WT) & \multicolumn{1}{c}{16:24(WT)} &  \multicolumn{2}{c}{}
\\ 
\hline
\multirow{4}{*}{PO} & $N_H$ (10$^{22}$ cm$^{-2}$) & 3.57$^a$ & 2.25$^a$ & \multicolumn{1}{c|}{2.78$^a$} & 2.63$^a$ & 2.85$^a$ & \multicolumn{2}{c}{}
\\ 
 & $\Gamma$ & 1.24$^a$ & 0.80$^a$ & \multicolumn{1}{c|}{0.99$^a$} & 1.35$^a$ & 1.45$^a$ & \multicolumn{2}{c}{}
\\
 & $\chi^2_{\nu}$/d.o.f. & 2.61/163 & 3.14/348 & \multicolumn{1}{c|}{2.67/221} & 2.61/123 & 2.89/90 & \multicolumn{2}{c}{}
\\
 & $F_{\rm{PO}}$ ($10^{-10}$ erg cm$^{-2}$ s$^{-1}$) & 4.0 & 7.0 & \multicolumn{1}{c|}{6.6} & 1.9 & 1.9 & \multicolumn{2}{c}{}
\\ 
\hline
 & $N_H$ (10$^{22}$ cm$^{-2}$) & 2.71$^{+0.41}_{-0.37}$ & 1.85$^{+0.23}_{-0.20}$ & \multicolumn{1}{c|}{2.14$^{+0.28}_{-0.27}$} & 1.35$^{+0.19}_{-0.18}$ & \multicolumn{1}{c}{2.17$^{+0.51}_{-0.45}$} &  \multicolumn{2}{c}{}
\\ 
 & $\Gamma$ & 8.33$^{+0.92}_{-0.90}$ & 7.77$\pm$0.53 & \multicolumn{1}{c|}{8.01$^{+0.73}_{-0.71}$} & 9.50$\pm$2.12 & \multicolumn{1}{c}{8.11$\pm$1.15} & \multicolumn{2}{c}{}
\\
 PO & $kT_{\rm{PC}}$ (keV) & 1.58$\pm$ 0.06 & 1.81$^{+0.04}_{-0.05}$ & \multicolumn{1}{c|}{1.72$^{+0.05}_{-0.06}$} & 1.47$^{+0.06}_{-0.04}$ & \multicolumn{1}{c}{1.32$^{+0.08}_{-0.07}$} &  \multicolumn{2}{c}{}
\\
 + & $R_{\rm{PC}}$ (km) & 0.87$^{+0.07}_{-0.06}$ & 0.98$\pm$0.04 & \multicolumn{1}{c|}{1.01$^{+0.06}_{-0.05}$} & 0.66$^{+0.04}_{-0.05}$ & \multicolumn{1}{c}{0.80$\pm$0.10} &  \multicolumn{2}{c}{}
\\
 BB & $\chi^2_{\nu}$/d.o.f. & 1.12/161 & 1.08/346 & \multicolumn{1}{c|}{0.95/219} & 1.02/121 & \multicolumn{1}{c}{1.54/88} & \multicolumn{2}{c}{}
\\
 & $F_{\rm{PO+BB}}$ ($10^{-10}$ erg cm$^{-2}$ s$^{-1}$) & 77.5 & 40.3 & \multicolumn{1}{c|}{61.5} & 11.7 & \multicolumn{1}{c}{25.1} & \multicolumn{2}{c}{}
\\
 & $P_{F-test}$ & 1.2$\times$10$^{-12}$ & 6.7$\times$10$^{-33}$ & \multicolumn{1}{c|}{1.1$\times$10$^{-20}$} & 1.3$\times$10$^{-11}$ & \multicolumn{1}{c}{3.4$\times$10$^{-5}$} &  \multicolumn{2}{c}{}
\\
\hline
\end{tabular}
}
\begin{small}
\begin{flushleft}
$^{a}$ The $\chi^2_{\nu}$ of the fit is larger than 2.0 and XSPEC can not show the uncertainty of the parameter.\\
\end{flushleft}
\end{small}
\end{table*}

\subsection{{\it Timing Analysis}}

The {\it Swift}/XRT observations listed in Table~\ref{Oblog} were performed with the PC and WT modes.
The data of the PC mode retain full imaging and spectroscopic resolution but the time resolution is limited to 2.5 s.
The data of the WT mode only cover the central 8$'$ of the field of view and one dimensional imaging is preserved with a better time resolution of 1.7 ms. 
The observations we used for analysis were divided into two groups (January and March) to get more counts for accurate temporal analysis.
We also applied solar system barycentric time correction with the task ``hdaxbary'' using JPL DE200 solar system ephemeris.
We restricted the data in the effective energy range (0.3--10 keV) of {\it Swift}/XRT and performed epoch-folding with the period centered at 405.3 sec \citep{Nase89,ISB75} using the resolution of 0.1 sec with 200 trials in the close neighborhood ($\sim$ 395--415 sec).
After getting the possible signal around 407.0 sec, we then checked the period with more delicate resolution (0.01 sec) in the range of  $\pm$1.0 sec (406--408 sec).
The light curve was folded with 32 bins and the epoch zero (54835.16972 MJD) was defined at the start of the good time interval (GTI) from the ObsID 1.
The most possible period in our detections is 407.69(2) sec at epoch 54837.01125 MJD with $\chi^2_{\nu}=152.0~\rm{for}~31~\rm{dof}$ and 407.26(1) sec at epoch 54899.95535 MJD with $\chi^2_{\nu}=52.6~\rm{for}~31~\rm{dof}$.
(The reference epoch was determined by the mid-point of the whole time span.)
The corresponding random probabilities of these two periods are close to 0. 
The uncertainty level of the period was determined from the $\chi^2$ value using eq.~\ref{eqno1} by \citet{Lea87}
\begin{equation}\label{eqno1}
  \dfrac{\sigma_P}{\Delta P} = 0.71(\chi_{\nu}^2-1)^{-0.63}
\end{equation}
where ${\Delta P=\dfrac{P^2}{T}}$ is the inferred Fourier spacing and $\nu$ is the degree of freedom.
We show the associated pulse profiles in Fig.~\ref{PsFile}.
We note that the normalized strength of the interpulse around phase 0.2 comparing to the main pulse around phase 0.5 increased significantly but the third peak around phase 0.8 appears to vanish entirely from January to March.

\subsection{{\it Spectral Analysis}}

Since the center of the source in all the PC mode observations is ``pile-up'' that leads to detector saturation, the central source photons has already been eliminated in the clean image.
In order to avoid the pronounced depression of counts in the center of our target, we therefore directly downloaded the spectral products of PC mode observations from the UK {\it Swift} Science Data Centre (www.swift.ac.uk/user$\_$objects/).
The WT mode observations were obtained from the High Energy Astrophysics Science Archive Research Center and proceeded to the spectral analysis with the standard data reduction.  
We downloaded the response matrices (rmf) from the official {\it Swift} website.
The ancillary files (arf) are generated by the task ``xrtmkarf'' with the {\it Swift}/XRT CALDB.
Because of a strong absorption feature around 0.5 keV, which is associated with the Oxygen edge \citep{Cus2006}, we only fitted our spectra in the 0.5--10 keV band using XSPEC 12.5.1 and rebinned the data with a minimum of 50 counts per bin to ensure $\chi^2$ statistics.
We produced a phase-averaged spectrum for each PC and WT observation with enough exposures to generate more than 30 degrees of freedom in statistics for spectral fit.
Among all the investigations, we applied a single power-law model to represent the main non-thermal emission of the accretion column/ outer gap (Table~\ref{NTspresult} \& Table~\ref{PSspresult}) and a single blackbody model to indicate the thermal emission contributed by the hot spot of the neutron star (Table~\ref{Tspresult} \& Table~\ref{PSspresult}).
If the $\chi_{\nu}^2$ of the result is larger than 1, an additional component is required to improve the fit.
An additional blackbody feature was also included in the composite model when we take the thermal surface emission of the neutron star (Table~\ref{NTspresult}) or from the accretion disk (Table~\ref{Tspresult}) into the consideration.
The best-fit spectral parameters of different models are all given with the errors at 90\% confidence level.

\begin{table*}
\caption{\small{Best--fit spectral parameters for the thermal spectroscopy of 1A 1118--615. PC and disk individually represent the related parameters of polar cap and accretion disk. The radius is measured from the normalization factor for a source distance of 4 kpc and the flux is measured in the range 0.5--10 keV. $P_{F-test}$ is the chance probability that the improvement of the fit, compared to the single blackbody model.}}\label{Tspresult}
\resizebox{2.15\columnwidth}{!} {
\begin{tabular}{cl|c|c|cc|cc|c} 
\hline
\multicolumn{2}{c|}{Observed Time} & Jan. 04 & Jan. 04 & Jan. 04 & Jan. 04 & Jan. 06 & Jan. 06 & Jan. 07
\\
\multicolumn{2}{c|}{/model} & 03:36(WT) & 10:05(WT) & 13:21(PC) & 13:21(WT) & 12:07(PC) & 12:07(WT) & 10:18(WT)
\\ 
\hline
\multirow{5}{*}{BB} & $N_H$ (10$^{22}$ cm$^{-2}$) & 0.84$^{+0.17}_{-0.14}$ & 0.47$^{+0.21}_{-0.17}$ & 0.49$^{+0.11}_{-0.10}$ & 0.36$^{+0.01}_{-0.02}$ & 1.10$^{+0.42}_{-0.38}$ & 0.97$^{+0.15}_{-0.14}$ & 0.29$\pm$0.03
\\ 
 & $kT_{\rm{PC}}$ (keV) & 1.77$\pm$0.09 & 2.27$^{+0.23}_{-0.19}$ & 2.19$^{+0.11}_{-0.10}$ & 2.01$^{+0.01}_{-0.02}$ & 2.65$^{+0.34}_{-0.28}$ & 1.89$\pm$0.09 & 2.05$^{+0.04}_{-0.03}$
\\
 & $R_{\rm{PC}}$ (km) & 1.21$\pm$0.09 & 0.83$^{+0.10}_{-0.09}$ & 0.85$^{+0.06}_{-0.05}$ & 1.15$^{+0.01}_{-0.02}$ & 0.52$^{+0.08}_{-0.07}$ & 0.98$^{+0.06}_{-0.07}$ & 1.25$^{+0.03}_{-0.02}$
\\
 & $\chi^2_{\nu}$/d.o.f. & 1.14/68 & 0.92/36 & 1.25/100 & 1.35/755 & 0.81/34 & 1.22/104 & 1.17/514
\\
 & $F_{\rm{BB}}$ ($10^{-10}$ erg cm$^{-2}$ s$^{-1}$) & 7.3 & 8.2 & 7.8 & 10.8 & 4.9 & 6.4 & 13.8
\\  
\hline
\multirow{8}{*}{2BB} & $N_H$ (10$^{22}$ cm$^{-2}$) & 1.29$^{+0.35}_{-0.31}$ & -- & 0.78$^{+0.42}_{-0.30}$ & 0.52$^{+0.07}_{-0.06}$ & -- & 1.22$^{+0.28}_{-0.22}$ & 0.36$^{+0.06}_{-0.04}$
\\ 
 & $kT_{\rm{PC}} (keV)$ & 1.66$^{+0.10}_{-0.09}$ & -- & 2.10$^{+0.14}_{-0.12}$ & 1.94$^{+0.03}_{-0.02}$ & -- & 1.82$^{+0.09}_{-0.10}$  & 2.01$\pm$0.04
\\
 & $R_{\rm{PC}}$ (km) & 1.35$^{+0.14}_{-0.12}$ & -- & 0.91$^{+0.09}_{-0.08}$ & 1.20$\pm$0.02 & -- & 1.04$^{+0.09}_{-0.08}$ & 1.29$^{+0.04}_{-0.03}$
\\
 & $kT_{\rm{disk}} (keV) $ & 0.14$^{+0.05}_{-0.03}$ & -- & 0.18$^{+0.09}_{-0.15}$ & 0.16$^{+0.02}_{-0.02}$ & -- & 0.12$^{+0.07}_{-0.03}$ & 0.10$^{+0.03}_{-0.02}$
\\
 & $R_{\rm{disk}}$ (km) & 106$^{+220}_{-75}$ & -- & 22$^{+14}_{-21}$ & 24$^{+9}_{-6}$ & -- & 121$^{+482}_{-101}$ & 109$^{+158}_{-66}$
\\
 & $\chi^2_{\nu}$/d.o.f. & 1.03/66 & -- & 1.24/98 & 1.27/753 & -- & 1.17/102 & 1.12/512
\\
 & $F_{\rm{2BB}}$ ($10^{-10}$ erg cm$^{-2}$ s$^{-1}$) & 9.5 & -- & 8.1 & 10.9 & -- & 7.3 & 13.9
\\
 & $P_{F-test}$ & 0.013 & -- & 0.24 & 4.6$\times$10$^{-11}$ & --  & 0.046 & 1.5$\times$10$^{-5}$
\\
\hline
\multicolumn{2}{c|}{Observed Time} & Mar. 04 & Mar. 08 & \multicolumn{1}{c|}{Mar. 08} & Mar. 14 & \multicolumn{1}{c}{Mar. 14} & \multicolumn{2}{c}{}
\\
\multicolumn{2}{c|}{/model} & 22:16(WT) & 04:47(WT) & \multicolumn{1}{c|}{15:51(WT)} & 00:39(WT) & \multicolumn{1}{c}{16:24(WT)} &  \multicolumn{2}{c}{}
\\
 \hline
\multirow{4}{*}{BB} & $N_H$ (10$^{22}$ cm$^{-2}$) & 1.43$^{+0.16}_{-0.15}$ & 0.76$^{+0.08}_{-0.07}$ & \multicolumn{1}{c|}{0.96$\pm$0.11} & 0.82$^{+0.14}_{-0.13}$ & 0.92$^{+0.18}_{-0.16}$ & \multicolumn{2}{c}{}
\\ 
 & $kT_{\rm{PC}} (keV)$ & 1.77$\pm$0.06 & 2.05$\pm$0.05 & \multicolumn{1}{c|}{1.94$^{+0.06}_{-0.05}$} & 1.59$\pm$0.06 & 1.52$^{+0.06}_{-0.07}$ & \multicolumn{2}{c}{}
\\
 & $R_{\rm{PC}}$ (km) & 0.70$\pm$0.04 & 0.79$^{+0.02}_{-0.03}$ & \multicolumn{1}{c|}{0.81$^{+0.03}_{-0.04}$} & 0.57$^{+0.03}_{-0.04}$ & 0.60$^{+0.05}_{-0.04}$ & \multicolumn{2}{c}{}
\\
 & $\chi^2_{\nu}$/d.o.f. & 1.55/163 & 1.65/348 & \multicolumn{1}{c|}{1.43/221} & 1.52/123 & 1.90/90 & \multicolumn{2}{c}{}
\\
 & $F_{\rm{BB}}$ ($10^{-10}$ erg cm$^{-2}$ s$^{-1}$) & 2.6 & 5.4 & \multicolumn{1}{c|}{4.8} & 1.2 & 1.1 & \multicolumn{2}{c}{}
\\  
\hline
\multirow{8}{*}{2BB} & $N_H$ (10$^{22}$ cm$^{-2}$) & 2.50$^{+0.34}_{-0.32}$ & 1.71$\pm$0.19 & \multicolumn{1}{c|}{1.94$^{+0.25}_{-0.23}$} & 1.21$\pm$0.17 & \multicolumn{1}{c}{2.06$^{+0.45}_{-0.42}$} &  \multicolumn{2}{c}{}
\\
& $kT_{\rm{PC}} (keV)$ & 1.59$^{+0.07}_{-0.05}$ & 1.82$^{+0.05}_{-0.04}$ & \multicolumn{1}{c|}{1.73$^{+0.06}_{-0.05}$} & 1.50$^{+0.05}_{-0.06}$ & \multicolumn{1}{c}{1.33$\pm$0.07} &  \multicolumn{2}{c}{}
\\
 & $R_{\rm{PC}}$ (km) & 0.85$\pm$0.06 & 0.96$\pm$0.04 & \multicolumn{1}{c|}{0.99$^{+0.05}_{-0.06}$} & 0.64$\pm$0.04 & \multicolumn{1}{c}{0.79$^{+0.09}_{-0.08}$} &  \multicolumn{2}{c}{}
\\
 & $kT_{\rm{disk}}$ (keV)& 0.14$\pm$0.02 & 0.14$\pm$0.01 & \multicolumn{1}{c|}{0.14$^{+0.01}_{-0.02}$} & 0.09$^{+0.01}_{-0.02}$ & \multicolumn{1}{c}{0.14$\pm$0.02} &  \multicolumn{2}{c}{}
\\
 & $R_{\rm{disk}}$ (km) & 179$^{+150}_{-81}$ & 152$^{+69}_{-44}$ & \multicolumn{1}{c|}{189$^{+119}_{-73}$} & 481$^{+749}_{-272}$ & \multicolumn{1}{c}{125$^{+136}_{-68}$} &  \multicolumn{2}{c}{}
\\
 & $\chi^2_{\nu}$/d.o.f. & 1.12/161 & 1.12/346 & \multicolumn{1}{c|}{0.96/219} & 1.05/121 & \multicolumn{1}{c}{1.52/88} &  \multicolumn{2}{c}{}
\\
 & $F_{\rm{2BB}}$ ($10^{-10}$ erg cm$^{-2}$ s$^{-1}$) & 6.6 & 8.6 & \multicolumn{1}{c|}{9.1} & 2.9 & \multicolumn{1}{c}{3.1} &  \multicolumn{2}{c}{}
\\
 & $P_{F-test}$ & 2.2$\times$10$^{-12}$ & 3.8$\times$10$^{-30}$ & \multicolumn{1}{c|}{3.6$\times$10$^{-20}$} & 8.1$\times$10$^{-11}$ & \multicolumn{1}{c}{1.9$\times$10$^{-5}$} &  \multicolumn{2}{c}{}
\\
\hline

\end{tabular}
}
\end{table*}
\begin{figure*}
\centering
\hspace*{\fill}\includegraphics[width=6.2cm,angle=-90]{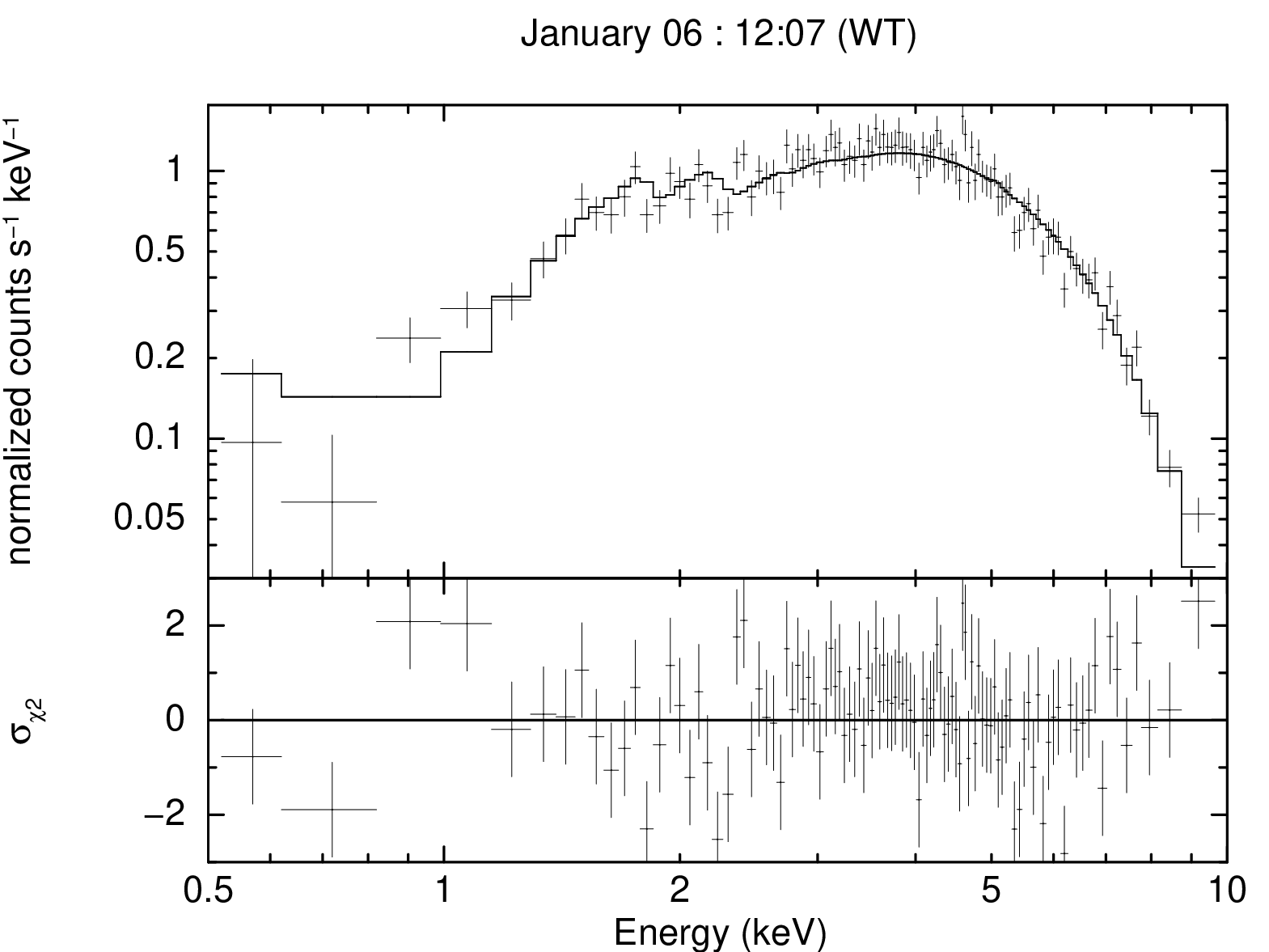}
\hspace*{\fill}\includegraphics[width=6.2cm,angle=-90]{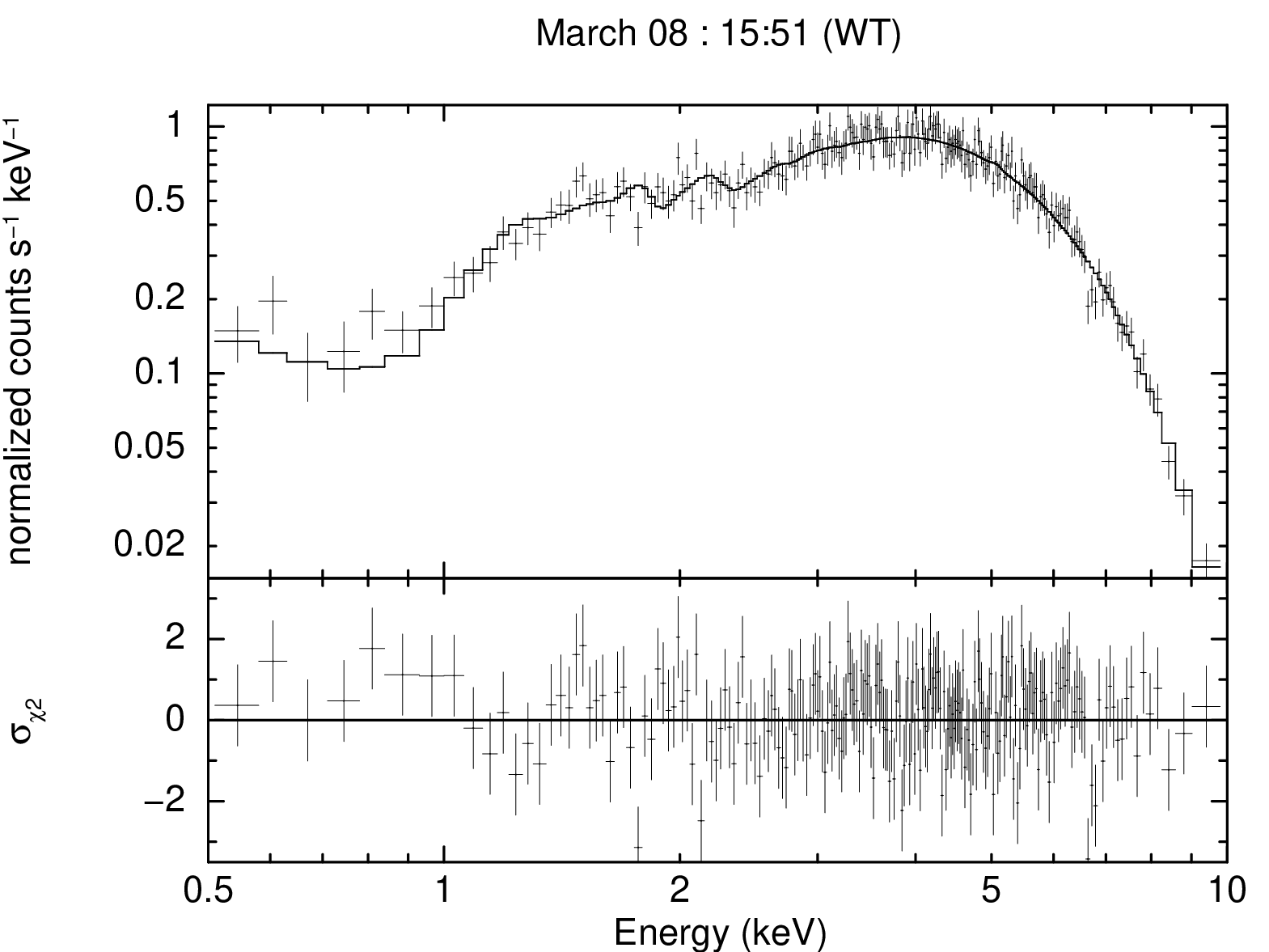}
\hspace*{\fill}
\caption{\small Spectral fits to blackbody models. (a).The left panel shows the fit to the single blackbody model for the WT mode data of ObsID 4 and (b).the right panel shows the fit to the composite blackbody model (BB+BB) for the WT mode data of ObsID 8. The corresponding null hypothesis probabilities of these two fits are all larger than 5\% and both of the results are attributed to be acceptable. The residuals in terms of sigmas with the error bars of size one are also shown.}
\label{accept_fit}
\end{figure*}
\begin{figure*}
\centering
\hspace*{\fill}\includegraphics[width=6.2cm,angle=-90]{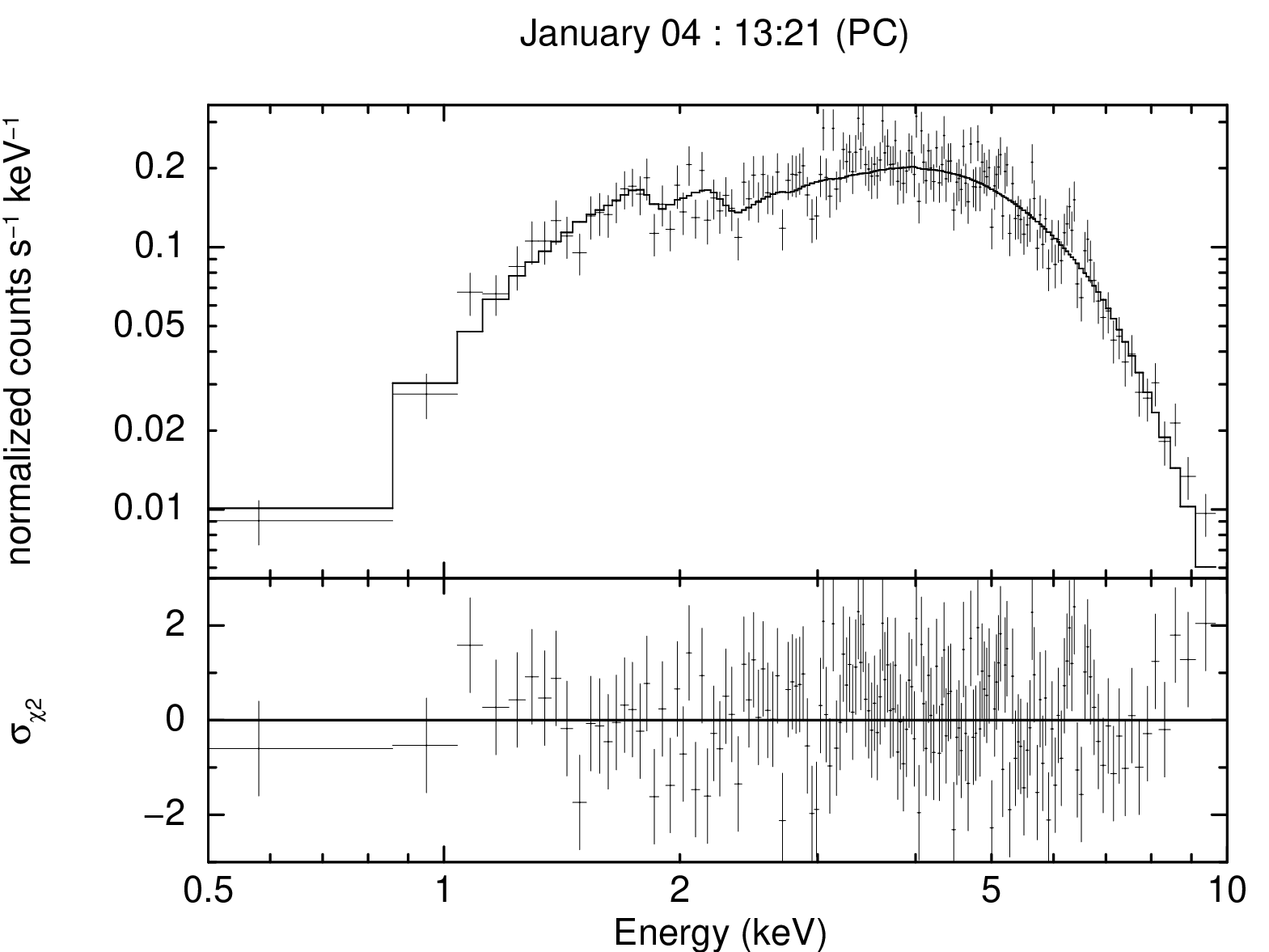}
\hspace*{\fill}\includegraphics[width=6.2cm,angle=-90]{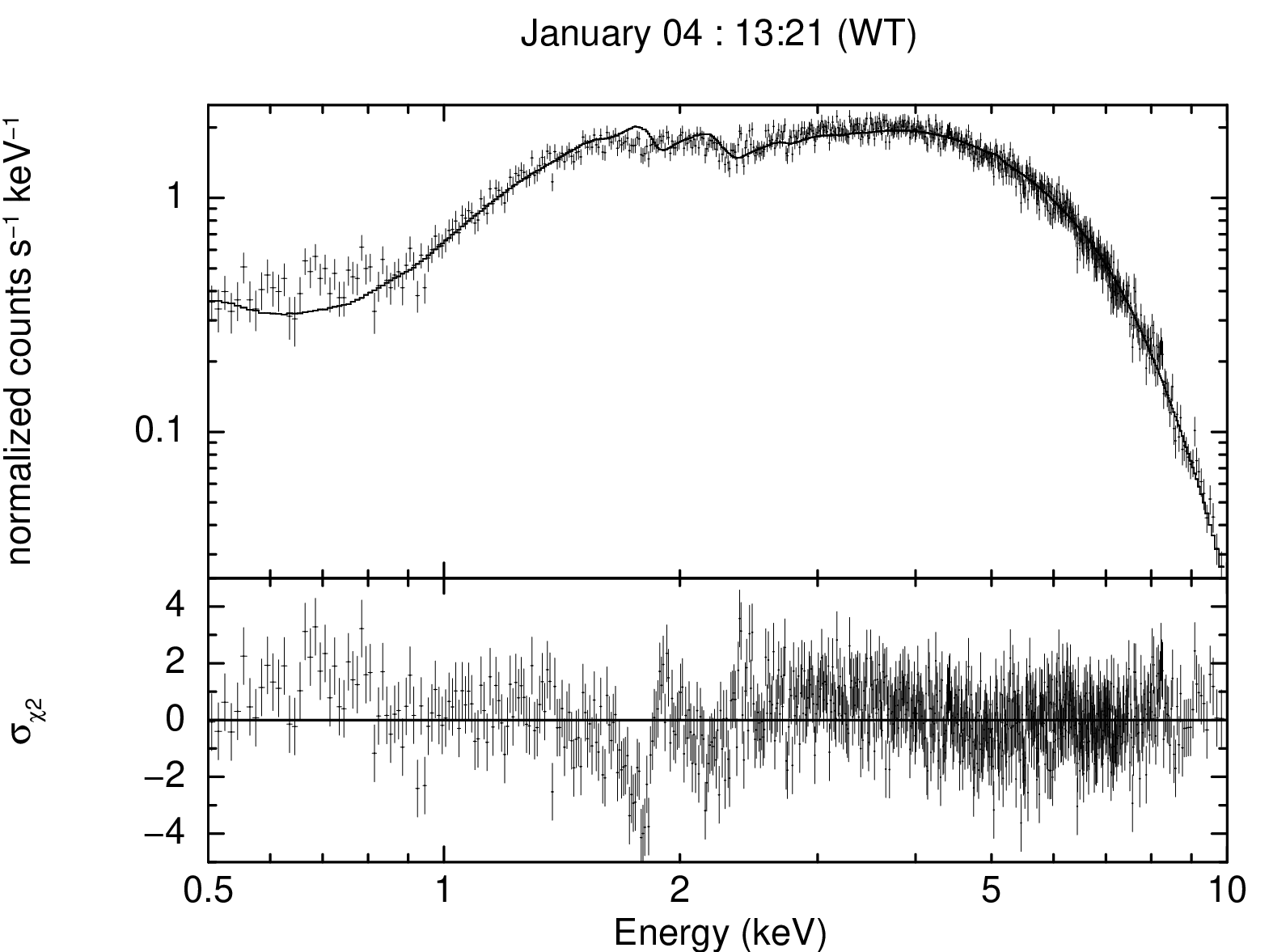}
\hspace*{\fill}\includegraphics[width=6.2cm,angle=-90]{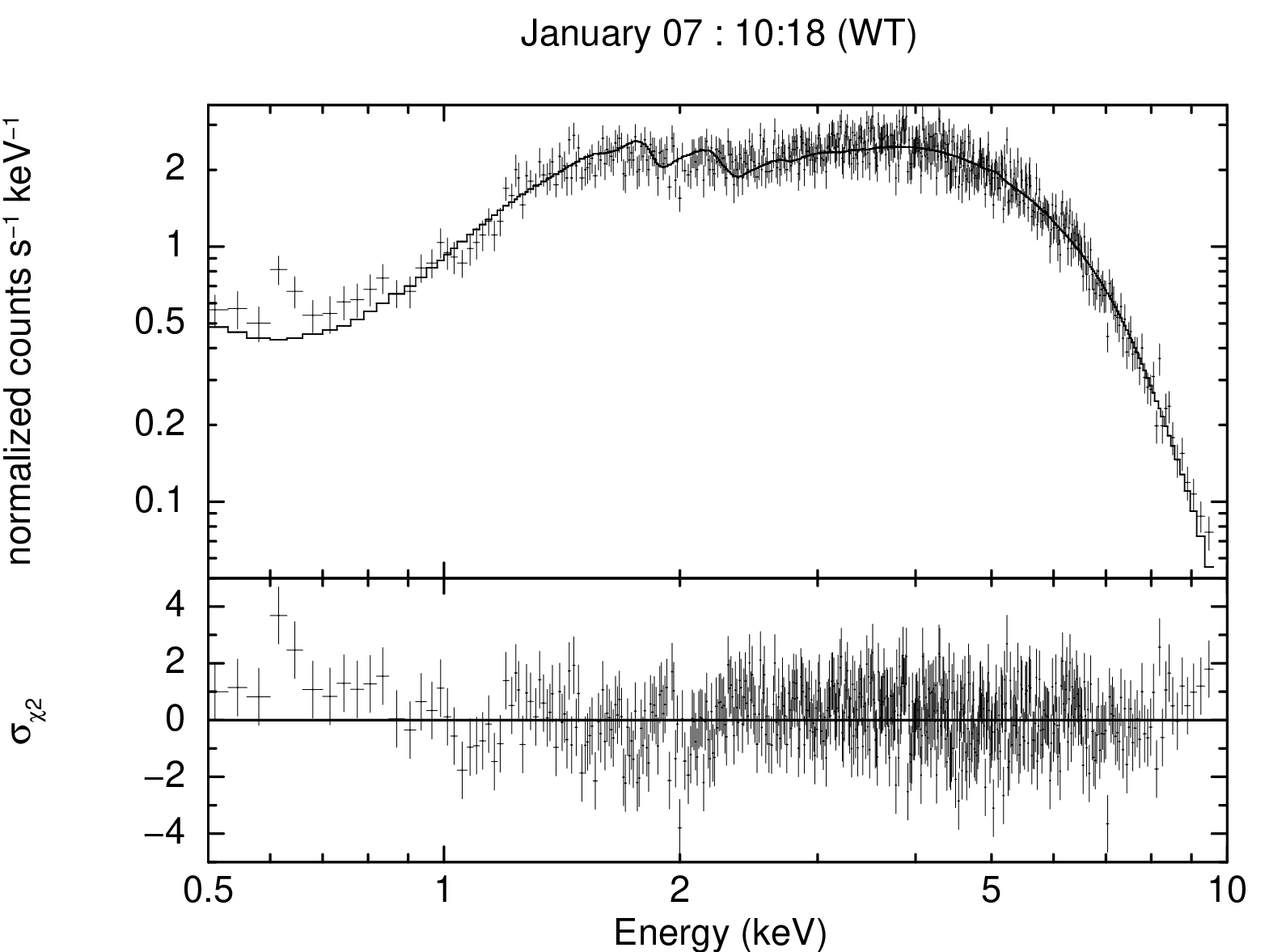}
\hspace*{\fill}\includegraphics[width=6.2cm,angle=-90]{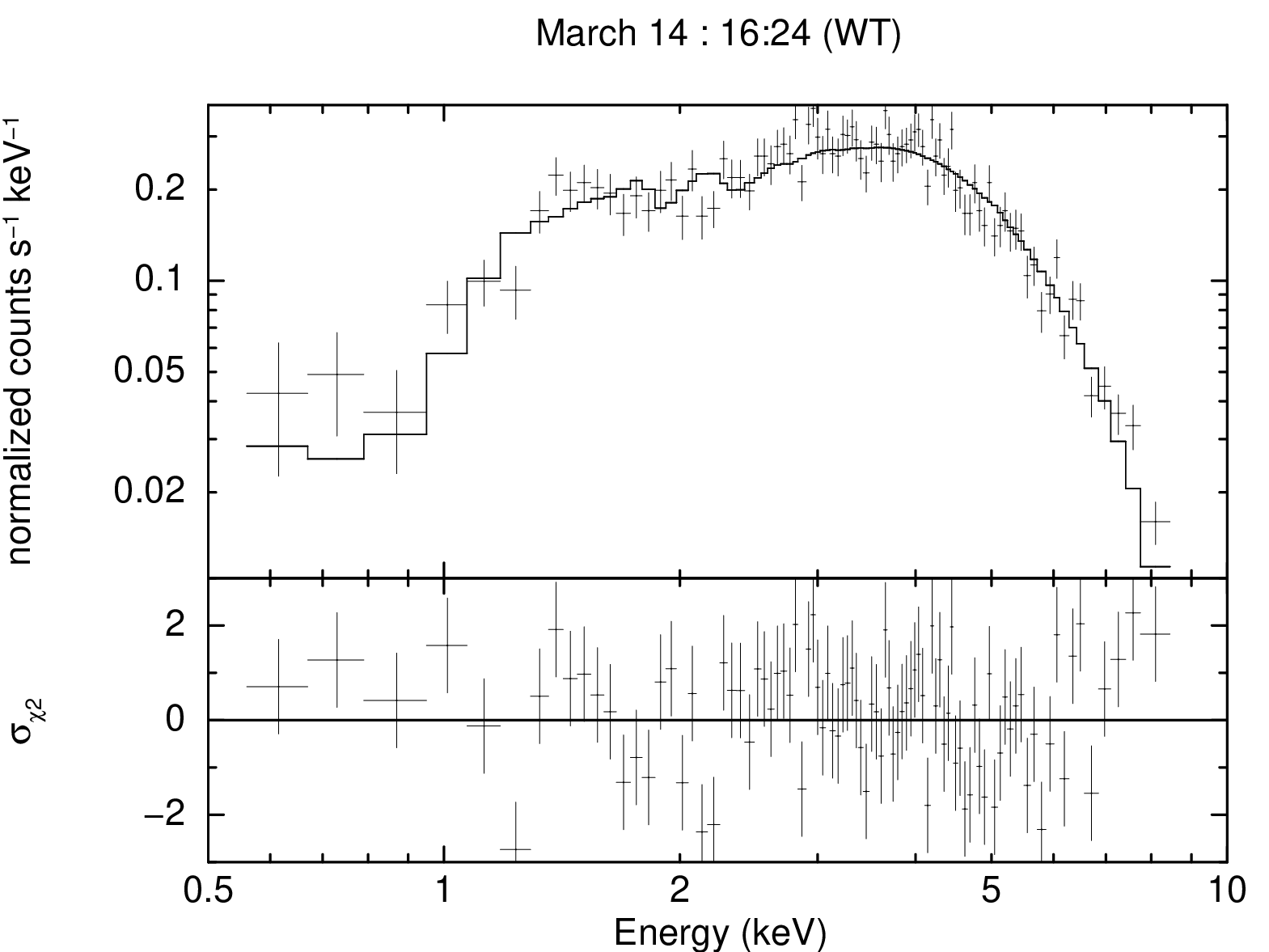}
\hspace*{\fill}
\caption{\small Spectral fits to blackbody models for 3 spectra observed in January and one spectrum observed in March. The January spectra are fitted to a single blackbody component; the March one is fitted to two blackbody components. All of these data have relatively long exposures than the other observations and the statistical results of these spectral fits are poor. Some strong absorbed features (e.g. 1.5 keV of Aluminum, 1.8 keV of Silicon et al.) caused by the uncertainty of the calibrations are clearly visible (the Oxygen edge at 0.5~keV had already been removed before performing the spectral analysis). There is also the strong instrumental Nickel contamination present in the high energy edge which sometimes is not fully subtracted. If all of these effects can be completely removed, the results of these fits will have significant improvements.}
\label{poor_fit}
\end{figure*}

For all observations, a single power-law model can not provide a reasonable fit as demonstrated in Table~\ref{NTspresult} (except for the observation of January 06).
On the other hand, a single blackbody model can provide an acceptable fit (Table~\ref{Tspresult} \& (a) of Fig.~\ref{accept_fit}) for the data observed in January when we took account the systematic uncertainties caused by calibrations  (Fig.~\ref{poor_fit}).
The spectra with long exposures seem to have worse statistics than those with short exposures.
This might be caused by the complicated absorption features associated with the response matrices \citep{Cus2006}.
A composite model with both power-law and blackbody can be fitted with the extracted spectra in March; however, the photon index ($\Gamma \sim 7.5-9$) is too soft compared with other X-ray binary pulsars (XBPs) and the inferred luminosity in the range 0.3--10~keV will exceed the Eddington luminosity.
A composite model can also fit most of the spectra in January, but the associated photon index of the power-law component is still too soft and the main contribution is dominated by the blackbody component.
We therefore ignore the contribution from the power-law for all the spectra and only considered a single/composite model with the blackbody component(s) to explain the spectral behavior. 

Two-blackbody components can improve the spectral fit during the outburst.
For ObsID 3 and 5 in WT mode, an F-test shows that the additional soft component during the outburst is significant at $> 99\%$.
But for the other 3 observations in January, an additional soft blackbody component does not improve the fits significantly.
In contrary, for the observations detected after the flaring in March, a two-blackbody model fits significantly better than a single blackbody one in Table~\ref{Tspresult} ((b) of Fig.~\ref{accept_fit}).
However, we note the WT spectra in ObsID 1, 2 \& 4 can be described by an acceptable single blackbody model with much shorter exposures than the other observations.  
It is unclear whether the single blackbody fits might be due to the larger statistical uncertainty in shorter exposures.
However, the single blackbody component fits with long exposures in January are still much better than the single blackbody component fits of March observations.
This indicates the variation of the spectral behavior might be real.      

We also note that the fitting parameters of Jan 04 observations (ObsID 1 \& 2) are very different.
This might be caused by the fact that the fitting parameter of the hydrogen absorption is very sensitive to the soft band spectrum if the the photon counts are low; a very small variation of the spectrum at soft band will result in a complete different hydrogen absorption, which will then cause a large variation in other fitting parameters. 
For the spectra of Jan 04 observations (ObsID 1 \& 2), if we fixed the hydrogen absorption to be $\sim 6.4\times10^{21}$~cm$^{-2}$, the ObsID 1\& 2 of the WT mode observations can both be fitted by a single blackbody model with $kT=1.86\pm 0.07$~keV, R~=$1.11\pm 0.05$~km with $\chi^2_{\nu}=1.21$ with 69 d.o.f and $kT=2.16^{+0.16}_{-0.13}$~keV, R~=$0.89\pm 0.07$~km with $\chi^2_{\nu}=0.95$ with 37 d.o.f. 
The variations of the parameters between these two spectra with the fixed hydrogen absorption are much smaller than those listed in the Table~\ref{Tspresult}. 
This indicates that the huge variations of the polar cap temperatures and radius within the same day may not be real. 

For all the spectra extracted in March, only the second spectrum of March 14 (ObsID 10) does not provide an acceptable fit with a two-blackbody model; this might be due to the serious Nickel contamination at 7--8~keV that is not fully subtracted from the calibration and many absorption features.
Some of the features are caused by the instrumental effects as shown by \citet{Cus2006}; however, the true spectra of the source might be much more complicated than our models and the instrumental effects probably do not explain all the poor fits to the data. 

\begin{table}
\caption{\small{Best--fit spectral parameters for the pulsed spectroscopy of 1A 1118--615. The radius is measured from the normalization factor for a source distance of 4 kpc and the flux is measured in the range 0.5--10 keV.}}\label{PSspresult}
\resizebox{1.00\columnwidth}{!} {
\begin{tabular}{cl|cc} 
\hline
\multicolumn{2}{c|}{Observed Time} & January & March 
\\
\multicolumn{2}{c|}{/model} &  (WT) & (WT)
\\ 
\hline
\multirow{4}{*}{PO} & $N_H$ (10$^{22}$ cm$^{-2}$) & 0.64 (fixed) & 1.88 (fixed) 
\\ 
 & $\Gamma$ & 0.23$\pm$0.07 & 1.04$^{+0.12}_{-0.13}$ 
\\
 & $\chi^2_{\nu}$/d.o.f. & 1.23/314 & 1.28/156
\\
 & $F_{\rm{PO}}$ ($10^{-10}$ erg cm$^{-2}$ s$^{-1}$) & 3.6 & 1.2
\\  
\hline
\multirow{5}{*}{BB} & $N_H$ (10$^{22}$ cm$^{-2}$) & 0.64 (fixed) & 1.88 (fixed)  
\\ 
 & $kT_{\rm{PC}}$ (keV) & 1.78$^{+0.10}_{-0.09}$ & 1.16$\pm$0.07 
\\
 & $R_{\rm{PC}}$ (km) & 0.70$\pm$0.05 & 0.84$^{+0.09}_{-0.08}$ 
\\
 & $\chi^2_{\nu}$/d.o.f. & 0.76/314 & 0.74/156 
\\
 & $F_{\rm{BB}}$ ($10^{-10}$ erg cm$^{-2}$ s$^{-1}$) & 2.8 & 0.83
\\  
\hline
\end{tabular}
}
\end{table}

\subsubsection{{\it Pulsed Spectral Analysis}}
  
To further study the emission from the neutron star, we also generated the pulsed spectrum of 1A 1118--615 with the data observed in the WT mode.
We divided each observation into pulsed and unpulsed emission depending on their pulsed phase shown in Fig.~\ref{PsFile}.
The light curve of 1A 1118--615 was folded with 32 bins and we assign the phase from 0.375 to 0.90625 (12th -- 29th of 32 bins) to be the pulsed phase of the 407.69 s period and the phase from 0.125 to 0.75 (4th -- 24th of 32 bins) to be the pulsed phase of the 407.26 s one.
We also used the same criterion to estimate all the pulsed photons in 0.3--10 keV for each data listed in Table~\ref{Oblog}.
To get enough counts for pulsed spectral fits, all the pulsed spectra in January and March were combined into a January spectrum and March one respectively.
The associated ancillary response files were also combined with weightings depending on the relative exposures (shown in Table~\ref{Oblog}).
The pulsed spectrum was then obtained by subtracting the unpulsed emission from the pulsed emission.
According to Table~\ref{Tspresult}, the average hydrogen absorption in the best fit to a single blackbody spectrum obtained at the outburst stage in January is $\sim 6.4\times10^{21}$~cm$^{-2}$ and is $\sim 1.9 \times 10^{22}$~cm$^{-2}$ to a two blackbody spectrum after the flare detection in March.
In the fits to the pulsed spectra, we fixed these values to associate with the pulsed-average spectra (Table~\ref{PSspresult}).
The difference of the absorption can be caused by the possibility that the absorber is related to the different amounts and/or the viewing angle of the material accreted by the pulsar along the line-of-sight. 

The pulsed spectra of January and March can not be well fitted by a single power-law, and we can not find any evidence of the non-thermal emission from each phase-averaged spectrum.
However, a single blackbody model can be applied to the pulsed spectra in January and March at 0.5--10~keV with $kT_{\rm{BB}}=1.78^{+0.10}_{-0.09}$~keV, $\rm{R}_{BB}=0.70\pm 0.05$~km with $\chi^2_{\nu}=0.76~\rm{for}~314~\rm{d.o.f}$ and $kT_{\rm{BB}}=1.16\pm 0.07$~keV, $\rm{R}_{BB}=0.84^{+0.09}_{-0.08}$~km with $\chi^2_{\nu}=0.74~\rm{for}~156~\rm{d.o.f}$.
The inferred pulsed flux is $(2.8\pm 0.6)\times 10^{-10}$~ergs cm$^{-2}$ s$^{-1}$ in January and $(8.3\pm 2.6)\times 10^{-11}$~ergs cm$^{-2}$ s$^{-1}$ in March.
The decrease of the luminosity might be caused by the cooling of the surface temperature.

\section{Discussion}

According to {\it Ariel V} \citep{ESWR75}, {\it CGRO}/BATSE \citep{Coe94} and {\it Swift}/BAT observations, the time separation of $\sim$ 17 years between the detection of each outburst for 1A 1118--615 suggests that the neutron star does not interact violently with the massive companion during most of the orbit \citep{VRPG99}.
If the outburst is only caused by the approaching of the neutron star to the periastron, the long time separations of the outbursts indicate a very flat elliptical orbit.

However, the orbital periods for known Be/XBPs range from several days (A~0538--66; \citealt{JGW80}) to hundred of days (X1145--619; \citealt{WWR81}) and no other convincing evidence can connect this long time separation to be the orbital period of 1A 1118--615.
The time separation of $\sim$ 1 month between the outburst of 2009 January detected by {\it Swift}/XRT and the flare of 2009 February detected by {\it INTEGRAL} might also indicate a possible orbital period.
One major concern is that we would expect to have observed much more flares during the interval of 17 years or at least more than one flares after the main outburst if the orbital period was only around 1 month. 
In addition, each main outburst detected every 17 years is the normal outbursts (type I, 10$^{36}$ -- 10$^{37}$ erg/s). 
However, we would expect to observe a giant burst for a long recurrence interval (type II, $> 10^{37}$ erg/s; \citealt{Cab2008}).  
On the other hand, if the orbital period for 1A~1118--615 is 17 years, this binary system will have the longest orbital period than all the other known Be/XBPs shown by \citet{Cor2009}.

We would expect to observe some evidences of the passage of the neutron star during the periastron in other wave bands. \citet{Coe94} found that the equivalent width of H$_{\alpha}$ varied on very short time scales around the outburst of 1992; this can be explained as the disruption of the circumstellar disk caused by the passage of the neutron star during the periastron.
This indicates that the assumption of the 17-year orbital period is consistent with the H$_{\alpha}$ equivalent width variation.
Further optical investigations and the study of high-energy emission mechanism of this Be/X-ray transient may help us to verify the orbital period and solve the related mystery.

\subsection{{\it Observational features detected by Swift}}

The pulsation in January was measured during the outburst while in March it was detected after the flare detection of February.
The pulsations of 1A~1118--615, shown in Fig.~\ref{PsFile}, obviously demonstrate that the pulsar has spun-up during the period between the outburst and flare events.
The pulse periods do not show variation during the different observations of January and March respectively.
As a result it is unclear whether the outburst or flare is responsible for the spinning-up of the pulsar.
But the change of the pulse profile for 1A 1118--615 may relate to the decrease of the pulsed luminosity; similar luminosity dependencies were also claimed for Cen X-3 \citep{Nase92}, LMC X-4 \citep{Lev91} and EXO 2030+375 \citep{PWS89}. 
In the evolution of the pulse profile for EXO 2030+375 following the first outburst decay, an interpulse appeared and became stronger as the luminosity decreased from the maximum and this trend continued until the main pulse had become entirely disappeared.
Then the profile became dominated by the contribution from the interpulse.
The same luminosity dependence was also evident for 1A~1118--615.
The pulse profiles of 1A 1118--615 show features similar to those of EXO 2030+375 \citep{PWS89}; the pulsed luminosity decreased and the normalized strength of the interpulse became significant relative to the main pulse after the flaring.
A long term survey is needed to further examine the dependence of the pulse profile and X-ray luminosity.

According to Table~\ref{Tspresult}, a single blackbody model can provide an acceptable fit with a temperature of $kT > 1$~keV and a small hot spot radius ($\sim 1$~km; assuming a source distance of 4 kpc; \citealt{CP85}) for most of the data observed in the outburst.
The thermal emission can be attributed to the ionized photons or the collision of the heated gas; however, the most plausible origin is from the surface (polar cap) emission of the neutron star \citep{LPM2007,LP2009}.
\citet{HNK2004} presented a clear physical picture to show the emitting process during the accretion.
In this picture, the radius of the accreting polar cap is $\sim 0.1R_{NS}$, where $R_{NS}$ is the radius of the neutron star, typically assumed to be 10 km.
In contrary to the spectra of most XBPs \citep{HNK2004}, we did not find any obvious evidence of a main non-thermal contribution.
Instead of X-rays emitted from the accretion column or outer gap, we detected strong emission from a surface layer.
This X-ray contribution during the outburst can be originated from the thermal emission around the compact source, which is similar to the polar cap emission of an isolated neutron star.

The second thermal feature ($R\geq 100~\rm{km}$ \& $kT < 0.2~\rm{keV}$) resembles the X-rays from the inner edge of an accretion disk \citep{HNK2004}.
This additional soft characteristic may become an important component when the pulsed flux of the source weakens.
Such X-ray thermal contributions are always observed in the bright XBPs with luminosity $>$~ 10$^{37-38}$~ergs/s \citep{Paul2002,Ram2002,MWAL2009}, but the source luminosity of 1A 1118--615 in 0.5 -- 10 keV observed by {\it Swift}/XRT is only $\sim (0.6-1.7)\times 10^{36}$~ergs~s$^{-1}$.
Our results might indicate that the soft excess contributed by the emission from the inner disk can generally be observed for the XBPs after flaring, not only for the bright targets. 
If the variation of the second thermal component is real, our spectral analyses in different time intervals may indicate that the surface (polar cap) emission of the neutron star dominates the observed X-rays from 1A 1118--615 at the beginning of outburst.
The fact that the thermal emission contributed by the accretion disk becomes significant at another specific time interval may represent that a different violent radiation process starts in one month and may be also associated with the flare detection.
Although the average source luminosity in different time intervals only changes by a small fraction, the cooling of temperature in the similar polar cap area clearly exhibit a decay of the surface emission from the neutron star.
The decrease of the pulsed emission derived from our phase-resolved spectrum also supports this viewpoint.

\subsection{\it Parameters of the binary system}
The parameters of the neutron star and the properties of the orbital motion for the binary system, 1A~1118--615, are not well understood. 
Furthermore, the possible orbital period of $\sim17$~yr is extremely large compared with 100~days for the typical Be/XBPs with long spin period. 
Therefore, it will be worth examining the parameters of the neutron star and orbital motion for further discussion on this unique high-mass X-ray binary system.  

The increasing X-ray emission at the outburst stage in January of 2009 allowed us to find a rotation period of $P\sim407.69(2)$ sec. 
On the other hand, a rotation period of $P\sim 409.2(8)$~sec was reported by {\it Chandra} observation on August 21 of 2003 in the quiescence stage of the source \citep{Rutl2007}. 
Assuming that this spin up of the neutron star from its quiescence stage is a result of a continuous accretion of matter onto the stellar surface of the neutron star, we can estimate the accretion rate and the magnetic field of the neutron star. 
The {\it Chandra} observation in 2003 detected an X-ray luminosity of $L_{\rm{X}}\sim 1.8\times 10^{34}~\mathrm{erg/s}$ in the 0.5-10~keV energy bands. 
Assuming the X-ray emission is from the accretion onto the stellar surface, the accretion rate at the surface is estimated as \citep{Camp98} 
\begin{eqnarray}\label{eqno2}
\dot{M}_{acc}&=&\frac{R_{NS}}{GM_{NS}}\eta_{BC}L_{X} \nonumber \\
&\sim& 5
\times 10^{13}\eta_{BC}M_{1.4}^{-1}R_6L_{X,34}~\mathrm{g/s} 
\end{eqnarray}
where $\eta_{BC}$ is the bolometric correction, $R_{NS}$ and $M_{NS}$ are stellar radius of and the mass of the neutron star, respectively. 
In addition, $L_{X,34}$ is the X-ray luminosity in units of $10^{34}~\mathrm{erg/s}$, and $R_6$ is the stellar radius of the neutron star in units of $10^6$~cm and $M_{1.4}$ is the mass of the neutron star in units of the solar mass.  

We assume that the accretion rate is constant in the quiescent stage before the onset of the outburst in 2009 January.  
Combining the estimated accretion rate $\dot{M}_{acc}$ with conservation
of the angular momentum that 
\begin{equation}\label{eqno4}
I\delta{\Omega}/\delta t=\dot{M}_{acc}\sqrt{GM_{NS}R_m}
\end{equation}
 with $R_m$ being the magnetospheric radius where the magnetic pressure is balanced with the pressure of the accretion disk, the strength of the magnetic field of the neutron star is given by
\begin{equation}\label{eqno4}
B\sim 2\times 10^{14}\eta_{BC}^{-3}M_{1.4}^{3/2}R_6^{-6} L_{X,34}^{-3}~\mathrm{G}
\end{equation}
where we used the momentum of inertia of $I=10^{45}~\mathrm{g\cdot cm^3}$.
The strength of the magnetic field can be estimated as $B\sim
2\times 10^{14}~\mathrm{G}$ for $M_{1.4}=1$ and $R=10^6$~cm.
This value is also similar to those extremely magnetized neutron stars (known as magnetar, \citealt{WT2006}) if the bolometric correction is $\eta_{BC}\sim 1$. 
On the other hand, a typical strength of the magnetic field of a young neutron star $B\sim 10^{12}~\mathrm{G}$ is expected if $\eta_{BC}\sim 5$. 
Accretion onto the neutron star surface is permitted when the magnetospheric radius $R_m$ becomes smaller than the corotation radius, $R_{cor}=(GM_{NS}P^2/4\pi^2)^{1/3}$, where $P$ is the period of the neutron star. 
This condition is satisfied when $\eta_{BC}\geq 2$.


It will be expected that the outburst occurs as a result of the interaction between the equatorial disk of the Be star and the neutron star at the periastron passage.
Near the periastron, the velocity of the neutron star, $v_{NS}$,  will
become $v_{NS}\la (GM_{Be}/R_{Be})^{1/2}(\sim 500~\mathrm{km/s})$, where
we used  stellar mass and radius of Be star of $M_{Be}=17M_{\odot}$ and
$R_{Be}=10R_{\odot}$, respectively \citep{SMJ2009}.
This implies that the velocity of the neutron star is in general larger than the typical velocity for the matter of the equatorial disk of the Be star, whose typical initial velocity is $<<1$~km/s \citep{MZW97,Port98}. 
The velocity of the neutron star dominates that of the matter of the equatorial disk of the Be star near the periastron. 
Then the capture of the matter from the equatorial disk of the Be star by the neutron star will occur at a radius of 
\begin{equation}\label{eqno5}
r_{acc}\sim 2GM_{NS}/v_{r}^2\sim 3.8\times
 10^{11}\left(\frac{M_{NS}}{1.4M_{\odot}}\right)\left(\frac{v_{NS}}
{100~\mathrm{km/s}}\right)^{-2}~cm 
\end{equation}
from the neutron star, where $v_r\sim v_{NS}$ is the velocity of the material of the equatorial disk relative to the neutron star.
 
At the outburst stage in January of 2009, the average X-ray emission at 0.5--10~keV was measured with a luminosity of $L_X\sim 1.5\times 10^{36}~\mathrm{erg/s}$, indicating that the accretion rate onto the
neutron star is 
\begin{equation}\label{eqno6}
\dot{M}_{acc}\sim 6.4\times 10^{-10} \dfrac{\eta_{BC}}{5} M_{\odot}/\mathrm{yr}.
\end{equation}
 This accretion rate for a young neutron star is about 2 orders of magnitude smaller than typical mass loss rate ($\dot{M}_w\sim
10^{-7}~M_{\odot}/\mathrm{yr}$; \citealt{Wat88}) of an equatorial disk around a Be star.  
We can also estimate the accreting density of the surrounding medium from Bondi accretion ($\dot{M_{acc}}=4\pi
r^2_{acc}\rho v_r$; where $\rho$ is the ambient density). 
Thus we obtain:
\begin{equation}\label{eqno7}
\rho=\dfrac{\dot{M_{acc}}}{4\pi r_{acc}^2 v_r} \sim 5\times 10^{-16} \eta_{BC}
\left(\frac{M_{NS}}{1.4M_{\odot}}\right)^{-2}\left(\frac{v_{NS}}
{100~\mathrm{km/s}}\right).
~\mathrm{g/cm^{3}} 
\end{equation}
This is several orders smaller than the typical value ($\rho\sim 10^{-10}-10^{-13}~\mathrm{g/cm^3}$) of the density at the inner region of the disk \citep{SMJ2009}, implying that (1) the neutron star is likely interacting with the outer edge of the Be star disk or (2) the disk of Be star in 1A~1118--615 is relatively thin.

Our spectral analyses show one thermal polar cap component during the outburst stage in January and show an additional thermal disk component after the flaring in March.
This might be explained as a result of the neutron star approaching the equatorial disk of the Be star.
When the neutron star first approached the equatorial disk, the accretion disk of the neutron star could have accumulated relatively little material, so the accretion disk would not have significant emission and we could only detect the polar cap emission of the neutron star.
The flare of February might be caused by disk instabilities of the accretion disk of the neutron star. 
This indicates that the neutron star would have accreted significant amount of material from the equatorial disk of the Be star during the periastron; this makes the thermal disk component visible in the March observations after the February flare.


\section{Conclusion}

The Be/X-ray transient 1A 1118--615 has been observed in outburst by {\it Swift} starting in January of 2009. 
Comparing with the epochs of two previous outbursts for 1A 1118--615 detected by {\it Ariel V} and {\it CGRO}/BATSE, the time separation of $\sim 17$~years might give a hint as the orbital period.
On the other hand, the time interval of $\sim 1$~month between the detection of the outburst and the flaring in 2009 might also give another indication of the orbital period; especially given the fact that the new Be/XBP (e.g. SAX~J2103.5+4545; \citealp{BSS2000}) was discovered to have a rapid orbital motion with a long spin period.  
We performed timing analysis on the {\it Swift}/XRT archive data and the pulsations at 407.69(2) sec and 407.26(1) sec were individually detected in January and March.
1A~1118--615 had obviously spun-up due to the accretion from the massive primary between the January and March observations.

We also investigated the X-ray spectral behavior during the outburst in January and after another flare detection in March. 
Almost all the phase-averaged spectra can be well-modeled by thermal emission.
During the outburst, we found that a single blackbody component with a higher temperature of $\sim 1$~keV and a small radius of $R\sim 1$~km provides a reasonable fit, and it indicates that most of the observed X-rays are from the polar cap emission of the compact star. 
After the flare detected by {\it INTEGRAL}, the following {\it Swift}/XRT observations display a significant soft thermal excess above the main surface emission from the neutron star in their spectral behavior.
The soft thermal excess shows a lower temperature of $\sim 0.1-0.2$~keV and a larger emitting radius of $\sim 100$~km; this component might be connected to the emission from the inner accretion disk of the binary system.
In contrary to other XBPs, we do not find a significant non-thermal component in the spectrum of 1A 1118--615.
Our results show that the thermal radiation from the hot spot of neutron star and from the reprocessing of the dense material in the inner disk provides the emission mechanism for 1A 1118--615 after the outburst.
The highest luminosity at 0.5--10~keV of this source detected after the outburst is only $\sim 2.7\times 10^{36}$ ergs/s. 
Our results also demonstrate that the soft excess contributed by the inner accretion disk may not only be observed in bright XBPs ($L_x \ga 10^{37-38}$~ergs/s).

\citet{HNK2004} gave a clear physical picture to explain the possible origin of the soft excess. 
Our discoveries based on the spectral analysis may provide a hint to examine the accreting phenomenon.
We also derived the physical properties of this accreting system using a simple model.
The accretion rate and the density of the circumstellar disk derived with the assumption of the elongated orbit of 1A~1118--615 are lower than those of the other general Be/XBPs.
This indicates the neutron star might pass the outer edge of the equatorial disk of the Be star. 

\section*{Acknowledgments}

The authors appreciate an anonymous referee for his/her fruitful comments.
We thank Dr. Kwong-Sang Cheng for the fruitful discussion to this manuscript.
This research has made use of the data obtained through the High Energy Astrophysics Science Archive Research Center Online
Service, provided by the NASA/Goddard Space Flight Center. 
This work was partially supported by the National Science Council through grants NSC 98-2811-M-008-044. 
CYH acknowledges support from the National Science Council through grants
NSC~96-2112-M-008-017-MY3 and NSC~95-2923-M-008-001-MY3. 
AKHK acknowledges support from the National Science Council through grants
NSC~96-2112-M-007-037-MY3.

\label{lastpage}

\end{document}